\documentclass[aps,pra,twocolumn,groupedaddress,nofootinbib,longbibliography]{revtex4-1}
\usepackage{amsmath}
\usepackage{amssymb}
\usepackage{graphicx}
\usepackage{mathrsfs}
\usepackage{ntheorem}
\usepackage{mathtools}
\usepackage{enumerate}
\usepackage{enumitem}
\usepackage[T1]{fontenc}
\usepackage{physics}
\usepackage{subfigure}
\usepackage{overpic}
\usepackage{epstopdf}
\usepackage{bm}
\usepackage{color,soul}
\usepackage[bookmarks=false]{hyperref}
\usepackage{xcolor}
\hypersetup{colorlinks=true,citecolor=blue,
linkcolor=blue,urlcolor=blue,pdfstartview=FitH,
bookmarksopen=true}

\newcommand{\I}{\mathrm{I}}

\begin{document}
\title{Visualizing quantum phase transitions in the XXZ model via the quantum steering ellipsoid}
\author{Ming-Ming Du}
\affiliation{Department of Physics, Shandong University, Jinan 250100, China}
\author{Da-Jian Zhang}
\email{zdj@sdu.edu.cn}
\affiliation{Department of Physics, Shandong University, Jinan 250100, China}
\author{Zhao-Yi Zhou}
\affiliation{Department of Physics, Shandong University, Jinan 250100, China}
\author{D. M. Tong}
\email{tdm@sdu.edu.cn}
\affiliation{Department of Physics, Shandong University, Jinan 250100, China}

\date{\today}
\begin{abstract}
The past two decades have witnessed a surge of interest in borrowing tools from quantum information theory to investigate quantum phase transitions (QPTs). The best known examples are entanglement measures whose nonanalyticities at critical points were tied to QPTs in a plethora of physical models. Here, focusing on the XXZ model, we show how QPTs can be revealed through the quantum steering ellipsoid (QSE), which is a geometric tool capable of characterizing both the strength and type of quantum correlations between two subsystems of a compound system. We find that the QSE associated with the XXZ model changes in shape with the QPTs, that is, it is a needle in the ferromagnetic phase, an oblate spheroid in the gapless phase, and a prolate spheroid in the antiferromagnetic phase. This finding offers an example demonstrating the intriguing possibility of unveiling QPTs in a geometrically visible fashion. Some connections between our results and previous ones are discussed.
\end{abstract}

\maketitle

\section{Introduction}

Understanding quantum phase transitions (QPTs) lies at the heart of quantum many-body physics \cite{Sachdev1999}. Traditionally, QPTs are dealt with the Landau-Ginzburg theory, which is built upon the notions of symmetry breaking, order parameter, and correlation length \cite{Sachdev1999}. Despite enormous success in relatively simple systems, the theory fails to capture QPTs exhibited in some complex systems. This can be either due to
the difficulty of identifying proper order parameters for
systems whose symmetry breaking patterns are unknown or because of the very absence of local order parameters, e.g., for QPTs involving topological order \cite{Wen1993,Read2000,Kitaev2006}. Currently, alternative approaches to characterizations of QPTs are under continuous development.

A line of such developments is to explore tools borrowed from quantum information theory to characterize QPTs, which has received a surge of interest over the past two decades \cite{2008Amico517}. The starting point of this line is the finding that  critical phenomena occurred in complex many-body systems have profound relations with bipartite entanglement of two subsystems \cite{2002Osterloh608}. This has motivated a series of studies that make use of entanglement measures like the entanglement of formation \cite{1998Wootters2245} and the entropy of entanglement \cite{1996Bennett2046} to quantitatively study QPTs for a plethora of physical models \cite{Osborne2002,Gu2003,Vidal2003,Refael2004,
Wu2004,Lambert2004,Gu2004,Gu2005,
Wei2005,Anfossi2005,Wu2006,Venuti2006,Oliveira2006,Buonsante2007,
Orus2008,Facchi2008,Son2009,
Rulli2010,Chiara2012,Hofmann2014,Sahling2015,Bayat2017,Pezze2017}. Meanwhile, further developments \cite{
Wolf2006,Legeza2006,
You2007,Legeza2007,Venuti2007,Zhou2008,
Rachel2012,
Cui2012,Sun2014,
Karpat2014,Radhakrishnan2016} have been carried out that establish the links between QPTs and various measures of other types of quantum correlations, such as quantum discord \cite{Dillenschneider2008,Maziero2010,Sarandy2009,Werlang2010,
Werlang2011,2012Ren60305,Sun2015,Piroli2017}, Bell nonlocality \cite{Deng2012,Justino2012}, and steered quantum coherence \cite{Hu2020,Xie2020}.

These measure-based proposals have proven to be useful in detecting QPTs for a large number of systems. However, it has been found that each measure has its limitations for different types of QPTs. For example, the concurrence was found to be unable to faithfully identify the second-order QPTs in some spin-$1/2$ models \cite{2004Vidal22107,Yang2005}, the Kosterlitz-Thouless QPT (KT-QPT) was shown to be undetectable by a number of measures \cite{Chen2016,Chen2008,Ye2020}, and the nonanalytic behavior of quantum discord was argued to be not completely reliable for detecting the critical points in some many-body systems \cite{Sun2010}.
Therefore, it deserves further effort to find more effective tools for characterizing QPTs.

In this work, we advocate the use of quantum steering ellipsoid (QSE)  \cite{Jevtic2014} for characterizing QPTs. Akin to the Bloch sphere which is a geometrical representation of the state space of a qubit, QSE is a faithful representation of an arbitrarily given state of two qubits in the three-dimensional (3D) Euclidean space, up to some local unitary transformations. The key advantage offered by this representation is that both the strength and type of quantum correlations are made manifest in terms of simple geometric entities like the shape and volume of QSE. Note that a measure of quantum correlations is only able to capture the strength of quantum correlations of a specific type, and different types of quantum correlations are described by different kinds of measures. We therefore anticipate that QSE is more effective in characterizing QPTs compared with a measure of quantum correlations adopted previously.

We examine the above idea by inspecting the QSE associated with the XXZ model and find that it changes in shape as well as in volume with the QPTs. As detailed below, in the course of varying the anisotropy parameter of the XXZ model, the shape of the QSE changes successively from a needle to an oblate spheroid and to a prolate spheroid. We show that these three types of shapes correspond to the ferromagnetic phase, the gapless phase, and the antiferromagnetic phase, respectively. This demonstrates that the QSE is capable of revealing QPTs in a geometrically visible fashion, which is, of course, not achievable in the previous measure-based proposals. Moreover, we find that the QSE happens to be a sphere at the KT phase transition point \cite{1note1}, indicating that the QSE is able to signal the KT-QPT, which cannot otherwise be detected by a number of measures as shown in Refs.~\cite{Chen2016,Chen2008,Ye2020}.

It should be mentioned that there are other proposals of making use of geometric entities to study QPTs, such as the ones based on the fidelity \cite{2006Zanardi31123,2007Zanardi2002,2007Cozzini14439,2008Zhou412001,2008Zhou492002,
2010Wang64301}, the ones based on the Berry phase \cite{2005Carollo157203,2006Zhu77206}, the ones based on the metric tensor \cite{2007Zanardi100603,2019Zhang42104}, and the one based on the quantum geometric tensor \cite{Venuti2007} (see the review paper \cite {2020Carollo1} for more information). Generally speaking, all of these proposals are unable to provide a geometrically visible characterization of QPTs as did in our proposal, for the obvious reason that the geometric entities adopted there are defined in a mathematically abstract way. Moreover, although having succeeded in signaling QPTs for various systems, these proposals do not provide much insight into the nature of quantum correlations involved in QPTs. This is contrary to our proposal as well as the aforementioned measure-based proposals.

To be self-contained, we organize the rest of this paper as follows. In  Sec.~\ref{sec2}, we give a brief introduction to QSE. In Sec.~\ref{sec3}, we describe the XXZ model. In Sec.~\ref{results}, we present our main results. In Sec.~\ref{discussions}, we discuss the connections between our results and previous ones. At last, we conclude this work with some necessary remarks in Sec.~\ref{conclusions}.

\section{quantum steering ellipsoid}\label{sec2}

Consider the bipartite situation that
Alice and Bob share a two-qubit state $\rho_{AB}$. In general, $\rho_{AB}$ can be expressed as
\begin{eqnarray}
\rho_{AB}=\frac{1}{4} \sum_{\mu, \nu=0}^3 \Theta_{\mu \nu} \sigma^{\mu}_{A} \otimes \sigma^{\nu}_{B},
\end{eqnarray}
where $\sigma_A^{\mu}$ ($\sigma_B^{\mu}$), $\mu=0,1,2,3$, stands for the Pauli matrix associated with the Alice's (Bob's) qubit, and $\Theta$ denotes a $4\times 4$ real matrix with elements $\Theta_{\mu \nu}=\tr(\rho_{AB} \sigma^{\mu}_{A}\otimes\sigma^{v}_{B})$. Note that $\Theta$ can be written in the form of a block matrix as follows:
\begin{eqnarray}
\Theta=\left( \begin{array}{cc}
1 & \bm{b}^T \\
 \bm{a} & T \\
 \end{array}
\right).
\end{eqnarray}
Here, $\bm{a}$ and $\bm{b}$ are the Bloch vectors of the reduced density matrices $\rho_{A}$ and $\rho_{B}$, respectively. That is, $\bm{a}=(a_1,a_2,a_3)^T$ with $a_\mu=\tr(\rho_A\sigma_A^\mu)$, and similar expressions hold for $\bm{b}$. $T$ is a $3\times 3$ real matrix describing the correlations.

Apparently, performing local measurements on Bob's side can change or steer the state of the qubit on Alice's side. The state that Alice's qubit can be steered to, referred to as the steered state hereafter, is generally different if the measurement performed is different. Upon taking into account all possible local measurements on Bob's side, Alice can get all the possible steered states on her side. It has been shown that the set of all the possible steered states can be geometrically represented as an ellipsoid in the 3D Euclidean space, i.e., the so-called QSE for Alice \cite{Jevtic2014},
\begin{eqnarray}\label{QSE}
\mathcal{E}_A=\left\{\frac{\bm{a}+T\bm{x}}{1+\bm{b}\cdot\bm{x}}|x\leq 1\right\},
\end{eqnarray}
where $\bm{x}$ is a 3D vector with $x$ denoting its length. Elements of $\mathcal{E}_A$ are simply the Bloch vectors of all the possible steered states. So, the QSE is an ellipsoid inside the Bloch sphere.

\begin{figure}
\includegraphics[width=0.35\textwidth]{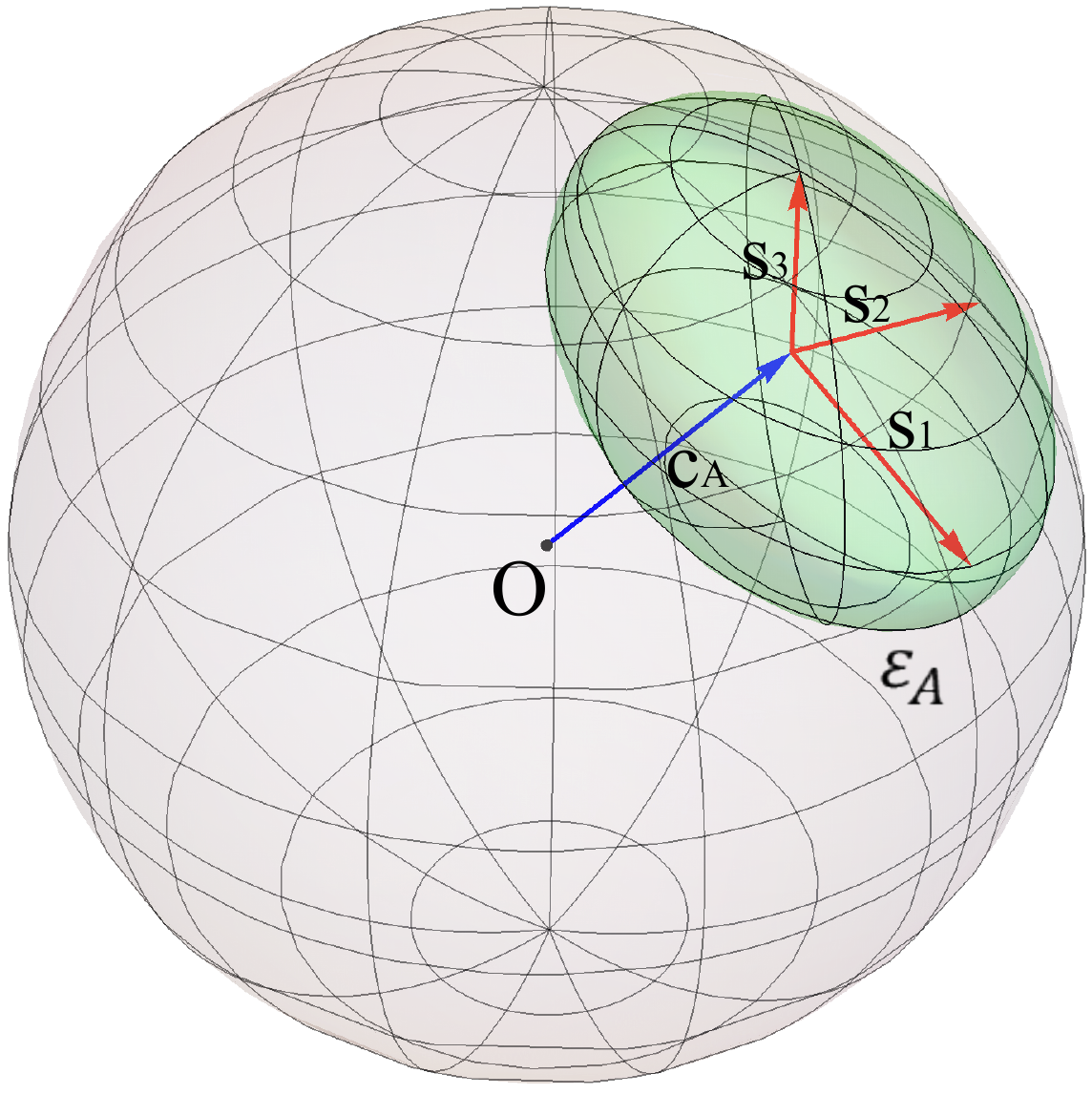}
\caption{Schematic of the quantum steering ellipsoid. The quantum steering ellipsoid $\mathcal{E}_A$ for Alice is an ellipsoid inside the Bloch sphere. It is determined by the ellipsoid center described by the blue vector $\bm{c}_A$ and the three semiaxes described by the red vectors $\bm{s}_i$.}
\label{fig1}
\end{figure}

The QSE is completely determined by the so-called ellipsoid center and  ellipsoid matrix. The ellipsoid center is described by
\begin{align}\label{center}
\bm{c}_A=\frac{\bm{a}-{T}\bm{b}}{1-b^2},
\end{align}
i.e., the vector pointing from the origin to the ellipsoid center, with $b$ denoting the length of $\bm{b}$. The ellipsoid matrix reads
\begin{align}\label{ellipsoid-matrix}
Q_{A}=\frac{1}{1-b^{2}}\left({T}-\bm{a} \bm{b}^{T}\right)\left(\I+\frac{\bm{b} \bm{b}^{{T}}}{1-b^{2}}\right)\left({T}^{{T}}-\bm{b} \bm{a}^{{T}}\right),
\end{align}
whose eigenvalues $q_i$ determine the lengths $s_i$ of the three semiaxes $\bm{s}_i$ of the ellipsoid as $s_i=\sqrt{q_i}$. All of the mentioned geometric entities like $\bm{c}_A$ and $\bm{s}_i$ are schematically shown in Fig.~\ref{fig1}. Note that Eqs.~(\ref{QSE}), (\ref{center}), and (\ref{ellipsoid-matrix}) are only well-defined if $b\neq 1$. If $b=1$, $\rho_{AB}$ is a product state and $\mathcal{E}_A$ is comprised of a single point determined by the Bloch vector of $\rho_A$.

The strength of quantum correlations may be described by the ellipsoid volume \cite{2017McCloskey12320}. Roughly speaking, the larger the ellipsoid volume is, the more quantum correlations $\rho_{AB}$ exhibits. In particular, if $\rho_{AB}$ is a product state, the QSE is a point and its volume is zero, and if $\rho_{AB}$ is one of the Bell states, the QSE is exactly the Bloch sphere and achieves its maximal volume over all possible two-qubit states. It has been shown that the volume can be expressed as \cite{Jevtic2014}
\begin{align}
V_A=\frac{64\pi}{3}\frac{|\det\rho_{AB}-\det\rho_{AB}^{{T}_{B}}|}{(1-b^2)^2},
\end{align}
where $T_B$ stands for the partial transpose with respect to Bob's qubit. This expression indicates that the ellipsoid volume $V_A$ is closely related to the strength of quantum correlations in view of the well-known Peres-Horodecki criterion \cite{1996Peres1413,1996Horodecki1}.

The relations between the QSE and different types of quantum correlations are rich and are still under continuous development \cite{Milne2014,Milne2015,Hu2015,Jevtic2015,
Hu2016,Nguyen2016a,Quan2016,Hu2016,Cheng2016,Nguyen2019,Zhang2019}. Here, we shall only mention a few fairly interesting results \cite{Jevtic2014}, which link different types of quantum correlations to different shapes as well as volumes of the QSE. It has been shown \cite{Jevtic2014} that $\rho_{AB}$ has zero discord for $A$ if and only if $\mathcal{E}_{A}$ degenerates to a radial line segment, and $\rho_{AB}$ has zero discord for $B$ if and only if $\mathcal{E}_{A}$ is 1D and $b=2\left|\bm{c}_{A}-\bm{a}\right| / l_{A}$, where $l_{A}$ is the length of $\mathcal{E}_{A}$. Moreover, $\rho_{AB}$ is separable if and only if $\mathcal{E}_{A}$ fits inside a tetrahedron that fits inside the Bloch sphere. On the other hand, any two-qubit state $\rho_{AB}$ with $V_A>4\pi/81$ must be entangled, but the converse may not be true, that is, an entangled state may have $V_A \leq 4\pi/81$.

Let us end the introduction to QSE by remarking that the QSE $\mathcal{E}_A$ for Alice is not identical to the QSE $\mathcal{E}_B$ for Bob in general. To obtain $\mathcal{E}_{B}$, we need only to make the substitutions $\bm{a} \rightarrow \bm{b}$, $\bm{b} \rightarrow \bm{a}$, and ${T} \rightarrow {T}^{{T}}$ in Eqs.~(\ref{center}) and (\ref{ellipsoid-matrix}) \cite{Jevtic2014}.

\section{spin-$1/2$ XXZ model}\label{sec3}

The XXZ model is an one-dimensional spin-$1/2$ chain in which the spins interact with each other through anisotropic Heisenberg interactions. Its Hamiltonian reads
\begin{eqnarray}\label{H-model}
H_\textrm{XXZ}=\sum_{j=1}^{N}S_{j}^{x} S_{j+1}^{x}+S_{j}^{y} S_{j+1}^{y}+\Delta S_{j}^{z} S_{j+1}^{z},
\end{eqnarray}
where $S_{j}^{\kappa}=\sigma_{j}^{\kappa} / 2$ $(\kappa=x, y, z), \sigma_{j}^{\kappa}$ are the Pauli spin-$1/2$ operators on site $j$, and $\Delta$ is the anisotropy parameter. Here, we have assumed the periodic boundary condition that $\sigma_{j+N}^{\kappa}=\sigma_{j}^{\kappa}$.
It is well-known that the model has three phases:\\
\begin{itemize}
\item $\Delta <-1$: The system is in a ferromagnetic phase, in which all the spins point in the same direction.
\item $-1<\Delta<1$: The system is in a gapless phase, in which the correlation decays polynomially.
\item $\Delta>1$: The system is in the antiferromagnetic phase.
\end{itemize}
QPTs occur at the boundaries between different phases. Specifically, a first-order QPT occurs at the critical point $\Delta=-1$, and the KT-QPT occurs at the critical point $\Delta=1$.

$H_\textrm{XXZ}$ cannot be diagonalized in general, but its energy spectrum can be obtained by employing Bethe ansatz. The Bethe ansatz solution gives the ground state energy $e_g$ as \cite{Yang1966,Yang1966a}
\begin{eqnarray}\label{energy}
&&e_{g}(\Delta)=\nonumber\\
&&\left\{\begin{array}{lc}
-\frac{\Delta}{4}, & \Delta \leqslant-1, \\
\frac{\Delta}{4}+\frac{\sin \pi \xi}{2 \pi} \int_{-\infty+\frac{i}{2}}^{\infty+\frac{i}{2}} d x \frac{1}{\sinh x} \frac{\cosh \xi x}{\sinh \xi x}, & -1<\Delta<1, \\
\frac{1}{4}-\ln 2, & \Delta=1,
\end{array}\right.
\end{eqnarray}
where $\Delta=\cos \pi \xi$. For $\Delta>1$, $e_g(\Delta)$ is obtained by changing $\xi=i \phi$ in Eq.~(\ref{energy}). On the other hand, it is worth noting that $H_\textrm{XXZ}$ respects some symmetries. Two symmetries of particular interest are the discrete parity $\mathbb{Z}_{2}$ symmetry over the plane $x y$, which transforms $\sigma_i^z$ into $-\sigma_i^z$ while keeps $\sigma_i^x$ and $\sigma_i^y$ unchanged, i.e.,
\begin{eqnarray}
\sigma_i^{x} \rightarrow \sigma_i^{x},~~\sigma_i^{y} \rightarrow \sigma_i^{y},~~ \sigma_i^{z} \rightarrow-\sigma_i^{z},
\end{eqnarray}
and the continuous $U(1)$ symmetry that rotates the spins in the $x y$ plane by any angle $\theta$, which transforms $\sigma_i^{\mu}$ as
\begin{eqnarray}
\sigma_i^{x} &\rightarrow& \cos\theta\sigma_i^{x}+\sin\theta\sigma_i^y,\\
\sigma_i^{y} &\rightarrow& \cos\theta\sigma_i^{y}-\sin\theta\sigma_i^x,\\
\sigma_i^{z} &\rightarrow&\sigma_i^{z}.
\end{eqnarray}
Besides, $H_\textrm{XXZ}$ respects some other symmetries like the translational  symmetry.

Analysis of the correlation functions $\expval{\sigma_i^\mu\sigma_j^\nu}$ for sites $i$ and $j$ can be simplified by taking the symmetries into account. It is easy to see that the $\mathbb{Z}_2$ symmetry implies that
\begin{eqnarray}\label{constrain1}
\langle\sigma_{i}^{z}\rangle=0,~~ \langle\sigma_{i}^{x} \sigma_{j}^{z}\rangle=0,~~ \langle\sigma_{i}^{y} \sigma_{j}^{z}\rangle=0,
\end{eqnarray}
where the state over which the expectation values are taken is assumed to be the ground state of the XXZ model. Here, we have assumed that there is no spontaneous symmetry breaking for the ground state under consideration. This assumption can be made justified for very small systems where the spontaneous symmetry breaking does not happen or by assuming that the ground state considered here is a superposition of the two degenerated ground states with opposite magnetization for the ferro and antiferromagnetic phases. The $U(1)$ symmetry implies that
\begin{eqnarray}\label{constrain2}
\langle\sigma_{i}^{x}\rangle=\langle\sigma_{i}^{y}\rangle=0,~~\langle\sigma_{i}^{x} \sigma_{j}^{x}\rangle=\langle\sigma_{i}^{y} \sigma_{j}^{y}\rangle,~~\langle\sigma_{i}^{x} \sigma_{j}^{y}\rangle=0.\nonumber\\
\end{eqnarray}
Taking into account all the symmetry constraints described by Eqs.~(\ref{constrain1}) and (\ref{constrain2}), we obtain that the reduced density matrix $\rho_{i,j}$ associated with sites $i$ and $j$ can be written as
\begin{align}\label{rhoij}
\rho_{i,j}=\left(
                       \begin{array}{cccc}
                         u & 0 & 0 & 0 \\
                         0 & w & y & 0 \\
                         0 & y & w & 0 \\
                         0 & 0 & 0 & u \\
                       \end{array}
                     \right),
\end{align}
where $u=(1+\langle\sigma_i^z\sigma_j^z\rangle)/4$, $w=(1-\langle\sigma_i^z\sigma_j^z\rangle)/4$, and {$y=(\langle\sigma_i^x\sigma_i^x\rangle + \langle\sigma_i^y\sigma_i^y\rangle)/4$}.

\section{Results}\label{results}

We show how the QPTs in the XXZ model can be revealed via QSE. Our basic idea is to signal the QPTs through the changes of the shape and volume of the QSE associated with two nearest-neighbor spins of the XXZ model. The reason why our idea shall work is as follows. It is well-known that QPTs are usually associated with some critical changes in the ground state energy of a many-body system due to quantum fluctuations. Moreover, it has been shown \cite{Wu2004} that for a Hamiltonian involving only nearest-neighbor interactions, the ground state energy can be obtained from the reduced density matrix of two nearest-neighbor constituents, say, $i$ and $i+1$; that is, the ground state energy is a function of the reduced density matrix $\rho_{i,i+1}$. So, the critical changes in the ground state energy generally lead to some critical changes in $\rho_{i,i+1}$, which shall lead to some critical changes in the shape and volume of the QSE associated with $\rho_{i,i+1}$ in general, since the QSE is a faithful representation of $\rho_{i,i+1}$ \cite{Jevtic2014}.

Now, following the above idea, we need to figure out the QSE associated with $\rho_{i,i+1}$. From Eq.~(\ref{rhoij}), it follows that
\begin{eqnarray}\label{ab-model}
\bm{a}=(0,0,0)^T,~~\bm{b}=(0,0,0)^T,
\end{eqnarray}
and
\begin{eqnarray}\label{T-model}
T=\begin{pmatrix}
    \expval{\sigma_i^x\sigma_{i+1}^x} & 0 & 0 \\
    0 & \expval{\sigma_i^y\sigma_{i+1}^y} & 0 \\
    0 & 0 & \expval{\sigma_i^z\sigma_{i+1}^z}
  \end{pmatrix}.
\end{eqnarray}
Apparently, $\bm{a}=\bm{b}$ and $T=T^T$, implying that the QSE for the $i$-th spin happens to be identical to that for the $(i+1)$-th spin. This simplifies our following analysis.
Substituting Eqs.~(\ref{ab-model}) and (\ref{T-model}) into Eqs.~(\ref{center}) and (\ref{ellipsoid-matrix}), we have
\begin{eqnarray}\label{center-model}
\bm{c}_A=(0,0,0)^T,
\end{eqnarray}
and
\begin{eqnarray}\label{Q-model}
Q_A=\begin{pmatrix}
    \expval{\sigma_i^x\sigma_{i+1}^x}^2 & 0 & 0 \\
    0 & \expval{\sigma_i^y\sigma_{i+1}^y}^2 & 0 \\
    0 & 0 & \expval{\sigma_i^z\sigma_{i+1}^z}^2
  \end{pmatrix}.
\end{eqnarray}
Here and henceforth, for the sake of notational consistency, we still adopt the notations like $c_A$, $Q_A$, and $V_A$ for the QSE for the $i$-th spin [or, equivalently, for the $(i+1)$-th spin]. We see from Eqs.~(\ref{center-model}) and (\ref{Q-model}) that the QSE is obtained from a unit sphere centered at the origin $(0,0,0)^T$ with its three semiaxes shrunk by the amounts of $\langle\sigma_{i}^{x} \sigma_{i+1}^{x}\rangle$, $\langle\sigma_{i}^{y} \sigma_{i+1}^{y}\rangle$, and $\langle\sigma_{i}^{z} \sigma_{i+1}^{z}\rangle$, respectively.  Note that it has been found that the correlation functions can be expressed as \cite{Shiroishi2005}
\begin{align}
&\langle\sigma_{i}^{x} \sigma_{i+1}^{x}\rangle=\langle\sigma_{i}^{y} \sigma_{i+1}^{y}\rangle=\frac{1}{2}\left[4 e_{g}(\Delta)-\Delta\langle\sigma_{i}^{z} \sigma_{i+1}^{z}\rangle\right],\label{cc1}\\
&\langle\sigma_{i}^{z} \sigma_{i+1}^{z}\rangle=4 \frac{\partial e_{g}(\Delta)}{\partial \Delta},\label{cc2}
\end{align}
with $e_g(\Delta)$ given by Eq.~(\ref{energy}). With the aid of Eqs.~(\ref{energy}), (\ref{cc1}), and (\ref{cc2}), we are able to work out any geometric property of the QSE, which is completely determined by Eqs.~(\ref{center-model}) and (\ref{Q-model}).

\begin{figure}
\centering
\includegraphics[width=0.45\textwidth]{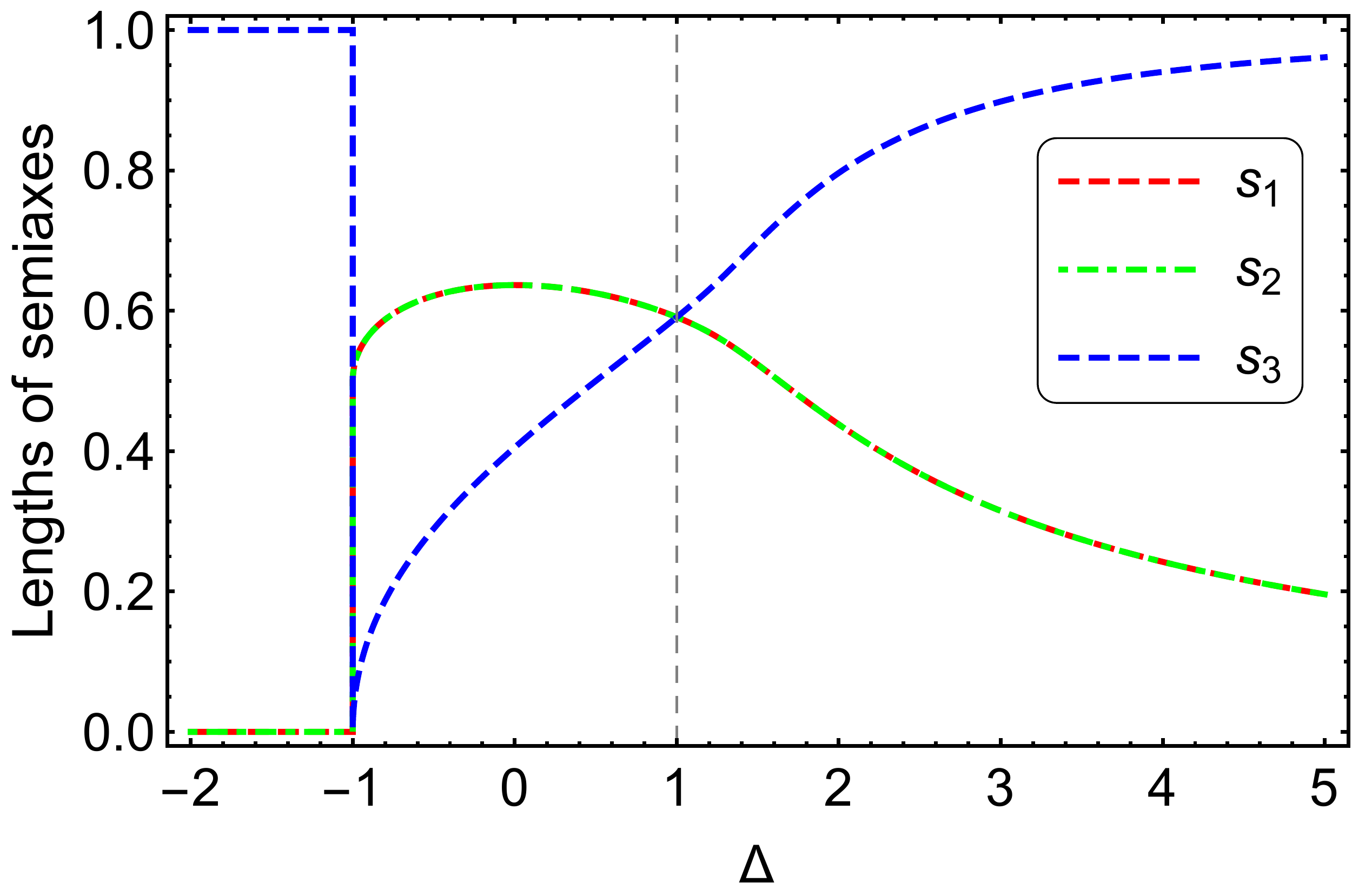}
\caption{Lengths of the three semiaxes of the quantum steering ellipsoid associated with $\rho_{i,i+1}$ as functions of the anisotropy parameter $\Delta$.}
\label{fig2}
\end{figure}

\begin{figure*}
\includegraphics[width=0.9\textwidth]{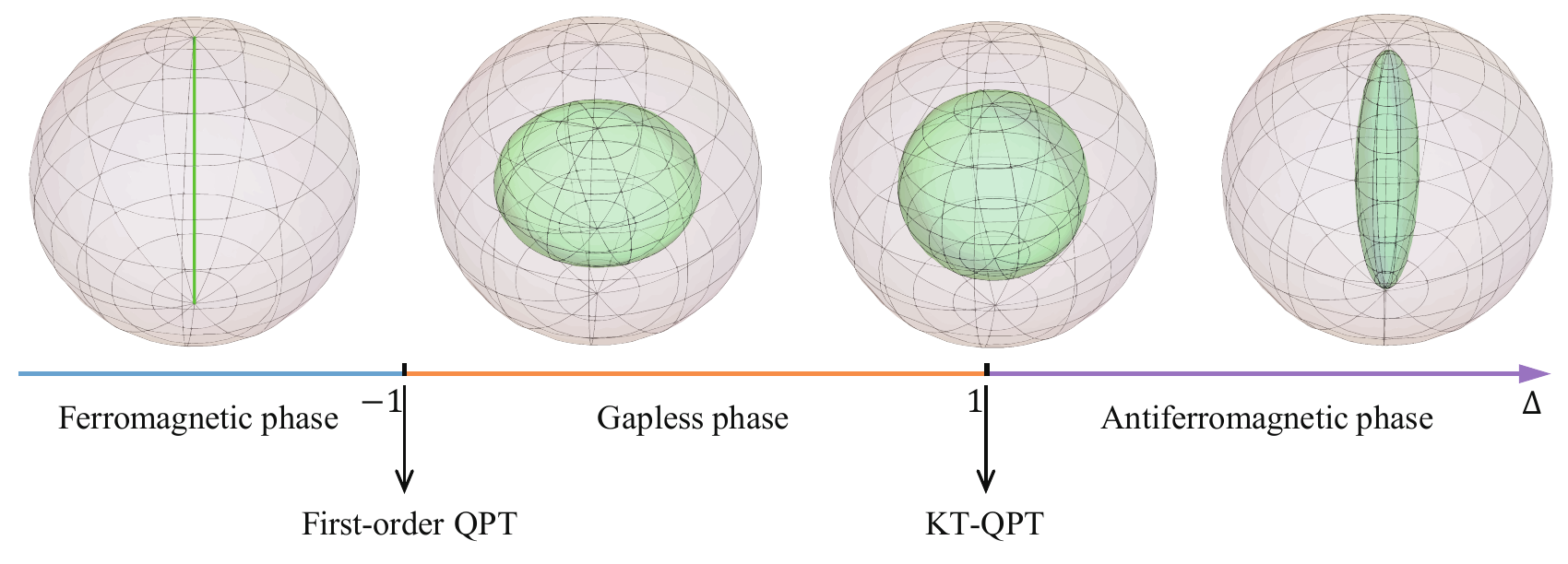}
\caption{Illustrations of different shapes of the quantum steering ellipsoid associated with $\rho_{i,i+1}$ for different phases of the XXZ model. The whole range of the values of $\Delta$ is divided into three regions which correspond to the ferromagnetic phase, the gapless phase, and the antiferromagnetic phase, respectively. The shape of the quantum steering ellipsoid changes with the quantum phase transitions. As indicated in the figure, it is a needle in the ferromagnetic phase, an oblate spheroid in the gapless phase, and a prolate spheroid in the antiferromagnetic phase. Moreover, the quantum steering ellipsoid changes abruptly in shape at the first-order quantum phase transition point $\Delta=-1$ and becomes a sphere at the Kosterlitz-Thouless quantum phase transition point $\Delta=1$. This establishes a connection between shapes of the quantum steering ellipsoid and different phases of the model, thereby demonstrating the intriguing possibility that quantum phase transitions can be revealed in a geometrically visible fashion.}
\label{fig3}
\end{figure*}

Figure \ref{fig2} shows the lengths of the three semiaxes $s_i$ of the {QSE} with $\rho_{i,i+1}$ as functions of $\Delta$. As can be seen from this figure, $s_1$ is equal to $s_2$ for all the values of $\Delta$. This follows immediately from the equalities connecting $s_i$'s to the correlation functions
\begin{eqnarray}\label{length-semiaxes}
s_1=|\langle\sigma_{i}^{x}  \sigma_{i+1}^{x}\rangle|, ~~
s_2=|\langle\sigma_{i}^{y}  \sigma_{i+1}^{y}\rangle|,~~
s_3=|\langle\sigma_{i}^{z} \sigma_{i+1}^{z}\rangle|,\nonumber\\
\end{eqnarray}
and Eq.~(\ref{cc1}). We observe from Fig.~\ref{fig2} that $s_i$'s abruptly change their values at $\Delta=-1$, which is the first-order QPT point. Another interesting observation is that the lengths of all the three semiaxes are equal at $\Delta=1$, which is the KT-QPT point. These two points $\Delta=\pm 1$ divide the whole range of the values of $\Delta$ into three regions: (i) $\Delta<-1$, for which $s_1=s_2=0$ and $s_3=1$; (ii) $-1<\Delta<1$, for which $s_1=s_2>s_3$; and (iii) $\Delta>1$, for which $s_1=s_2<s_3$. It is worth noting that these three regions correspond to the three phases listed in the previous section.

Let us inspect the shape of the QSE in these three regions. In the first region which corresponds to the ferromagnetic phase, the QSE is a needle since $s_1=s_2=0$ but $s_3>0$. This is pictorially shown in Fig.~\ref{fig3} (see the first Bloch sphere there which contains a green vertical line). The QSE changes in shape abruptly at the first order QPT point $\Delta=-1$. When entering into the second region which corresponds to the gapless phase, the QSE becomes an oblate spheroid (see the second Bloch sphere in Fig.~\ref{fig3} containing an oblate spheroid colored in green). As the value of $\Delta$ increases from $-1$ to $1$, we observe from Fig.~\ref{fig2} that the difference between the length of $s_1$ ($s_2$) and that of $s_3$ is increasingly small and eventually becomes zero at $\Delta=1$. This indicates that the QSE is increasingly closer in shape to a sphere and eventually becomes one at the KT-QPT point (see the third Bloch sphere in Fig.~\ref{fig3} containing a sphere colored in green). Keeping on increasing $\Delta$, we finally reach the third region which corresponds to the antiferromagnetic phase. Here, the QSE is a prolate spheroid as $s_1=s_2<s_3$ (see the fourth Bloch sphere in Fig.~\ref{fig3} containing a spheroid colored in green). Now, it is clear that the QSE changes in shape with the QPTs. This demonstrates the intriguing possibility that QPTs can be revealed in a geometrically visible fashion.

\begin{figure}
\includegraphics[width=0.45\textwidth]{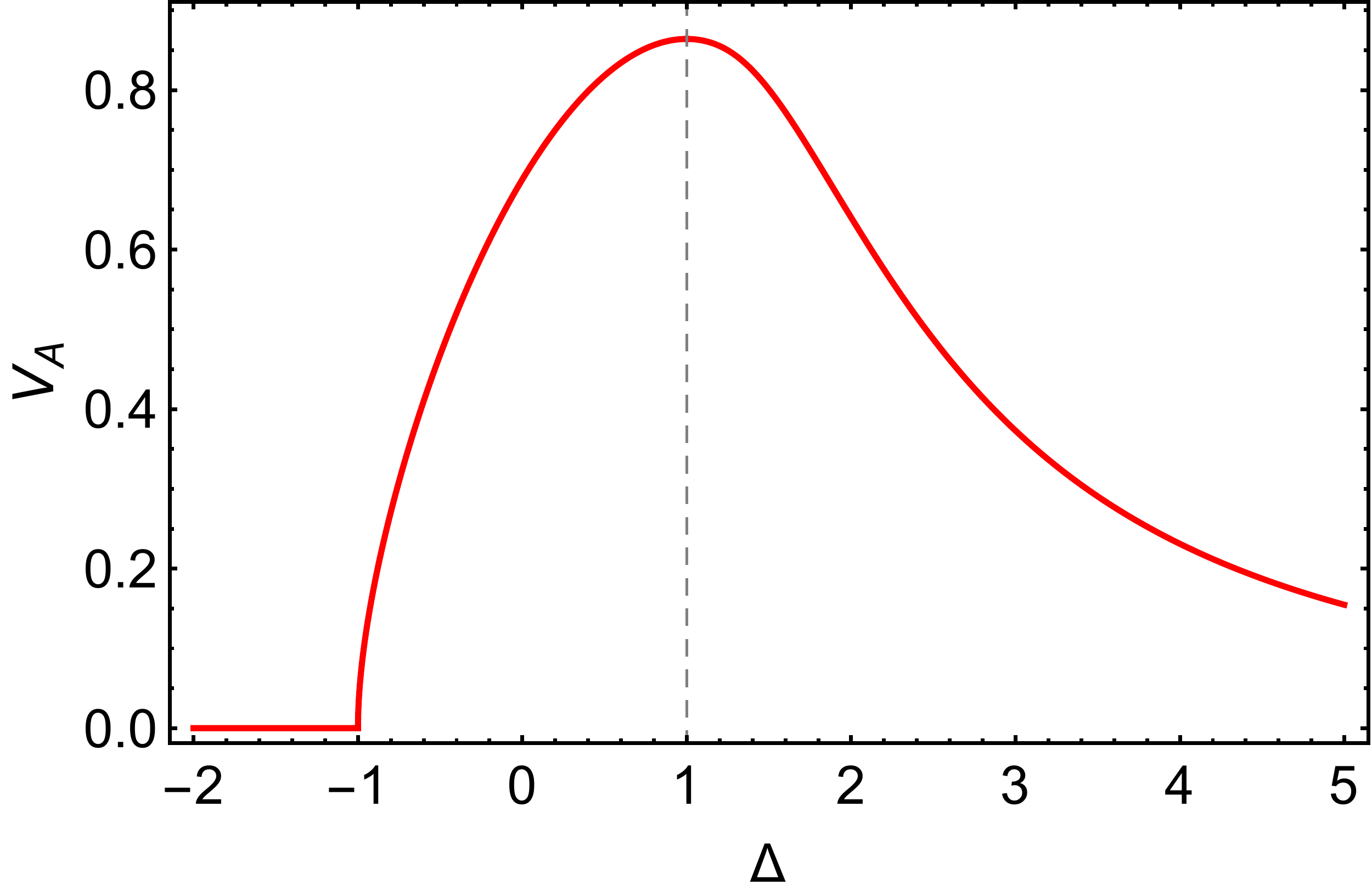}
\caption{Volume $V_A$ of the quantum steering ellipsoid associated with $\rho_{i,i+1}$ as a function of the anisotropy parameter $\Delta$. $V_A$ is zero in the first region $\Delta<-1$. It changes from zero to be larger than zero at the first-order quantum phase transition point $\Delta=-1$, indicating the appearance of quantum correlations. Keeping on increasing $\Delta$, we see that $V_A$ first increases rapidly in the second region $-1<\Delta<1$ and then decreases slowly in the third region $\Delta>1$. Its maximum is attained at the Kosterlitz-Thouless quantum phase transition point $\Delta=1$. The changes in the value of $V_A$ reflects the changes in the strength of quantum correlations as detailed in the text.}
\label{fig4}
\end{figure}

To gain further insight into the connection between the QSE and the QPTs, we numerically figure out the volume of the QSE in Fig.~\ref{fig4}. We observe that the volume $V_A$ of the QSE is zero in the first region corresponding to the ferromagnetic phase. This is simply because the QSE is a radial line in this region and therefore the state $\rho_{i,i+1}$ exhibits zero quantum discord, i.e., there is no quantum correlation. At the first-order QPT point $\Delta=-1$, the value of $V_A$ turns from zero to be strictly larger than zero, indicating the appearance of quantum correlations for $\Delta>-1$. When entering the second region, which corresponds to the gapless phase, we see that $V_A$ increases as $\Delta$ increases and attains its maximal value at the KT-QPT point $\Delta=1$. This indicates that quantum correlations become increasingly strong in the course of increasing $\Delta$ in this region. By contrast, when $\Delta>1$, corresponding to the antiferromagnetic phase, $V_A$ decreases as $\Delta$ increases. So, the strength of quantum correlations becomes increasingly small in the course of increasing $\Delta$
in this region.

\section{Discussions}\label{discussions}

So far, we have presented our main finding, which can be thought of as a geometrically visible picture that depicts the connection between the QPTs in the XXZ model and the shapes as well as volumes of the QSE. Note that the QPTs in the XXZ model have been alternatively studied with the aid of a number of measures of quantum correlations in the previous works. In the following, we show that the connection presented here sheds some new lights on these measure-based works. We would like to take the concurrence \cite{Yang2005}, quantum discord \cite{Dillenschneider2008,Sarandy2009}, Bell inequality \cite{Justino2012} and {difference of bond strength \cite{Luo2019} as four examples} to illustrate the above point.

\textit{Example 1: the concurrence \cite{Yang2005}.} The concurrence is an entanglement measure defined as \cite{1998Wootters2245}
\begin{eqnarray}
\mathcal{C}(\rho_{i,i+1})=\max\{0,\sqrt{\lambda_1}-
\sqrt{\lambda_2}-\sqrt{\lambda_3}-\sqrt{\lambda_4}\},
\end{eqnarray}
where $\lambda_i$'s are the eigenvalues of $\rho_{i,i+1}\rho_{i,i+1}^\prime$ arranged in descending order, with $\rho_{i,i+1}^\prime=\sigma^y\otimes\sigma^y\rho_{i,i+1}^*\sigma^y\otimes\sigma^y$ being the time-reversed density matrix. It has been found that the concurrence $\mathcal{C}(\rho_{i,i+1})$ displays some nonanalytical behaviors at $\Delta=-1$ and therefore can be used to signal the first-order QPT of the XXZ model \cite{Yang2005}. Using Eq.~(\ref{length-semiaxes}) and the results in Refs.~\cite{Gu2003,Gu2005,Yang2005}, which express $\mathcal{C}(\rho_{i,i+1})$ in terms of the correlation functions, we have that
\begin{eqnarray}\label{concurrence-QSE}
\mathcal{C}(\rho_{i,i+1})=
\begin{cases}
  0, & \Delta\leq -1, \\
  \frac{s_1+s_2+s_3-1}{2}, & \Delta> -1.
\end{cases}
\end{eqnarray}
Equation (\ref{concurrence-QSE}) establishes a connection between the concurrence and the lengths $s_i$ of the three semiaxes of the QSE. Note that $s_i$ abruptly changes its values at $\Delta=-1$, which can be seen from Fig.~\ref{fig2}. This means that the derivative of $s_i$ with respect to $\Delta$ is divergent at $\Delta=-1$, which immediately leads to the finding that $\mathcal{C}(\rho_{i,i+1})$ is nonanalytical at $\Delta=-1$.

\textit{Example 2: quantum discord \cite{Dillenschneider2008,Sarandy2009}.} Quantum discord $\mathcal{Q}$ is a measure of nonclassical correlations between two subsystems of a quantum system \cite{Ollivier2001}. Explicitly, it can be expressed as
\begin{align}\label{QD}
\mathcal{Q}(\rho_{i,i+1})
=\min _{\mathcal{M}} \sum_{k} p_{k} S\left(\rho_i^{k}\right)+S\left(\rho_{i+1}\right)-S\left(\rho_{i,i+1}\right).
\end{align}
Here, $S(\cdot)$ denotes the von Neumann entropy. $\mathcal{M}=\left\{M_{k}\right\}$ is a POVM measurement performed on the $(i+1)$-th qubit, with $M_{k} \geqslant 0$ and $\sum_{k} M_{k}=I$. $p_k$ denotes the probability of getting outcome $k$, and $\rho_i^k$ represents the post-measurement state of the $i$-th qubit. It has been found \cite{Dillenschneider2008,Sarandy2009} that $\mathcal{Q}(\rho_{i,i+1})$ is nonanalytical at $\Delta=1$ and therefore can be used to signal the KT-QPT. Note that both $S(\rho_{i+1})$ and $S(\rho_{i,i+1})$ are smooth functions at $\Delta=1$. So, nonanalyticities of $\mathcal{Q}(\rho_{i,i+1})$ stem from the term $\min _{\mathcal{M}} \sum_{k} p_{k} S\left(\rho_i^{k}\right)$. Using the explicit expression of $\min _{\mathcal{M}} \sum_{k} p_{k} S\left(\rho_i^{k}\right)$ for a Bell diagonal state \cite{Shi2011}, we obtain that
\begin{align}\label{smin}
\min _{\mathcal{M}} \sum_{k} p_{k} S\left(\rho_i^{k}\right)=\min \left\{h\left( s_{1}\right), h\left( s_{2}\right), h\left( s_{3}\right)\right\},
\end{align}
with
\begin{align}
h(g)=-\frac{1+g}{2} \log_2 \frac{1+g}{2}-\frac{1-g}{2} \log_2 \frac{1-g}{2},
\end{align}
where $g \in[0,1]$. Noting that $h(g)$ is a monotonically decreasing function of $g$, we have that the term $\min _{\mathcal{M}} \sum_{k} p_{k} S\left(\rho_i^{k}\right)$ is determined by the longest axis of the QSE. As can be seen from Fig.~\ref{fig2}, the longest axis is ${s}_1$ (or ${s}_2$) on the left side of $\Delta=1$ but is ${s}_3$ on the right side of $\Delta=1$. It follows immediately that the term $\min _{\mathcal{M}} \sum_{k} p_{k} S\left(\rho_i^{k}\right)$ is nonanalytical at $\Delta=1$, which further leads to the finding that $\mathcal{Q}(\rho_{i,i+1})$ is also nonanalytical at $\Delta=1$.

\textit{Example 3: Bell inequality \cite{Justino2012}.} A measure of the Bell nonlocality can be defined as the maximal violation of the CHSH inequality,
\begin{eqnarray}
\mathcal{B}(\rho_{i,i+1})=\max_{\bm{m}_i,\bm{n}_j}\abs{\expval{B_{CHSH}}},
\end{eqnarray}
where $\expval{B_{CHSH}}=\expval{A_1\otimes B_1}+\expval{A_1\otimes B_2}+\expval{A_2\otimes B_1}-\expval{A_2\otimes B_2}$, with $A_i=\bm{m}_i\cdot\bm{\sigma}$ and $B_j=\bm{n}_j\cdot\bm{\sigma}$. It has been found \cite{Justino2012} that the measure  $\mathcal{B}(\rho_{i,i+1})$ displays some nonanalytical behaviors at $\Delta=\pm 1$ and therefore can be used to signal both the first-order QPT and the KT-QPT. Using Eq.~(\ref{length-semiaxes}) and the results in Ref.~\cite{Justino2012}, which express $\mathcal{B}(\rho_{i,i+1})$ in terms of the correlation functions, we have that
\begin{align}\label{B}
\mathcal{B}(\rho_{i,i+1})=2\max\left\{\sqrt{s_1^2+s_2^2},\sqrt{s_1^2+s_3^2},\sqrt{s_2^2+s_3^2}\right\}.
\end{align}
Apparently, $\mathcal{B}(\rho_{i,i+1})$ is determined by the longest and second longest axes of the QSE. As can be seen from Fig.~\ref{fig2}, these two axes change from $\{s_1,s_3\}$ (or equivalently $\{s_2,s_3\}$) when $\Delta<-1$ to $\{s_1,s_2\}$ when $-1<\Delta<1$ and back to $\{s_1,s_3\}$ (or equivalently $\{s_2,s_3\}$) when $\Delta>1$. That is, one of these two axes changes at $\Delta=\pm 1$. This leads to the finding that $\mathcal{B}(\rho_{i,i+1})$ is nonanalytical at $\Delta=\pm 1$.

{\textit{Example 4: the difference of bond strength \cite{Luo2019}.} The difference of bond strength was introduced to signal first-order QPTs for many-body systems with Hamiltonians of the form $H(\lambda)=H_0+\lambda H_I$, where $\lambda$ denotes a driving parameter. Apparently, due to the linear dependence of $H(\lambda)$ on $\lambda$, the ground state energy varies linearly with $\lambda$ and can be expressed as $e_g(\lambda)=e_0+\lambda e_I$. Associated with a first-order QPT, a jump of $e_I$ will appear at the phase transition point $\lambda_t$. This reflects the structural change of the ground state. It is also expected that a similar jump of $e_0$ appears at $\lambda_t$ and may eliminate the singularity of $e_g(\lambda)$ caused by the jump of $e_I$. To avoid the elimination and therefore detect $\lambda_t$, the difference of bond strength is defined as
\begin{align}\label{D}
\mathcal{D}=e_0-\mathrm{sgn}(\lambda_t)e_I,
\end{align}
where the minus sign reflects the spirit of the bond reversal method. Using Eq.~(\ref{length-semiaxes}) and the results in Ref.~\cite{Luo2019}, which express $\mathcal{D}$ in terms of the correlation functions, we have that around the first-order phase transition point $\Delta=-1$, $\mathcal{D}$ can be expressed as
\begin{align}\label{DBS-QSE}
\mathcal{D}=
\begin{cases}
  -\frac{1}{4}(s_1+s_2-s_3), & \mbox{if } \Delta<-1; \\
  -\frac{1}{4}(s_1+s_2+s_3), & \mbox{if } -1\leq\Delta <0.
\end{cases}
\end{align}
Equation (\ref{DBS-QSE}) establishes a connection between the difference of bond strength and the lengths $s_i$ of the QSE. Note that $s_i$ abruptly changes its values at $\Delta=-1$, which can be seen from Fig.~\ref{fig2}. This immediately leads to the finding that the difference of bond strength $\mathcal{D}$ exhibits a jump at $\Delta=-1$.
 }

\section{Concluding remarks}\label{conclusions}

We have shown how the QPTs in the XXZ model can be revealed through the associated QSE. The basic idea is that, since QPTs are tied to critical changes of quantum correlations in general, they shall manifest themselves in terms of geometric entities of QSE. Assisted by the relations between quantum correlations and QSE found earlier \cite{Jevtic2014}, we have developed this idea into an explicit connection between different phases in the XXZ model and different types of shapes of the associated QSE as shown in Fig.~\ref{fig3}.

According to the connection established here, a QPT in the model occurs whenever the associated QSE changes its shape. Specifically, the QSE changes from a needle to an oblate and to a prolate spheroid when $\Delta$ passes through the first-order QPT point and the KT-QPT point successively. This demonstrates that QPTs can be revealed in a geometrically visible fashion, which, to our knowledge, has never been achieved before. Aided by the established connection, we have shed some lights on a number of previous works, showing that the nonanalyticities of the measures considered there are rooted in the semiaxes of the QSE.

While the present work focuses on the specific XXZ model, we anticipate that the idea of exploring QSE to signal QPTs can find its applications in a more general physical context \cite{2note2}. This is because QSE is a faithful representation of quantum correlations and therefore any critical change in quantum correlations shall lead to some critical change in QSE. A general framework for connecting QPTs to QSE is beyond the scope of this work but is highly desirable for the following reason. Apart from the advantage of demanding no \textit{a prior} knowledge of order parameters, the framework is expected to be universally effective in detecting QPTs, since QSE is capable of characterizing both the strength and type of quantum correlations.

We finally remark on the experimental detection of QPTs with QSE.
It is well-known that QPTs are assumed to occur at the absolute zero temperature, which is unattainable experimentally because of the third law of thermodynamics. Hence, in practice, one must work at a small temperature as close as possible to the absolute zero temperature, in order to detect QPTs \cite{Werlang2010,Werlang2011,Karpat2014,Chen2016}. However, at a small but nonzero temperature, there may exist a gap between theory and experiment so that a well-established theory for detecting QPTs may not work in experiments. Fortunately, the validation of QSE has been verified experimentally in Ref.~\cite{Zhang2019}. This indicates that the connection between QPTs and QSE established here for the XXZ model as well as the general connection to be explored in a future study is likely to hold in experiments.

\begin{acknowledgments}
We acknowledge support from the National Natural Science Foundation of China through Grant No. 11775129.
\end{acknowledgments}


\begin{thebibliography}{93}
\bibitem [{1no()}]{1note1}%
  \BibitemOpen
  \href@noop {} {}\bibinfo {note} {{The finding that the QSE is a sphere at
  the KT phase transition point is a consequence of the accidental SU(2)
  symmetry when $\Delta=1$. It is therefore an interesting result specific to
  the XXZ model and not a general signature of KT-QPTs.}}\BibitemShut {Stop}%
\bibitem [{2no()}]{2note2}%
  \BibitemOpen
  \href@noop {} {}\bibinfo {note} {{Based on our experience, we
  would like to point out that while QSE is generally capable of unveiling
  QPTs, different types of QPTs are tied to different characters of
  QSE.}}\BibitemShut {Stop}%
\end{thebibliography}
%

\end{document}